\documentclass[12pt]{article}
\usepackage{amsfonts}
\usepackage{amsmath}
\usepackage{amssymb,amsbsy}
\usepackage{bm}
\usepackage{makeidx}
\usepackage{latexsym}
\usepackage{graphicx}
\usepackage{epsfig}
\usepackage{subfigure}
\usepackage{dsfont}
\usepackage{mathcomp}
\usepackage{color}
\usepackage{mathrsfs}
\usepackage{epsf}
\usepackage{authblk}
\newcommand{\beq}{\begin{eqnarray}}% can be used as {equation} or  {eqnarray}
\newcommand{\eeq}{\end{eqnarray}}
\newcommand{\be}{\begin{equation}}% can be used as {equation} or  {eqnarray}
\newcommand{\ee}{\end{equation}}

\newcommand{\gapp}{\mathrel{\raise.3ex\hbox{$>$}\mkern-14mu
              \lower0.6ex\hbox{$\sim$}}}
\newcommand{\lapp}{\mathrel{\raise.3ex\hbox{$<$}\mkern-14mu
              \lower0.6ex\hbox{$\sim$}}}



\newcommand{\preprintno}[1]
{\vspace{-2cm}{\normalsize\begin{flushright}#1\end{flushright}}\vspace{1cm}}

\title{\preprintno{{\bf KCL-PH-TH/2012-025}}Quantifying Astrophysical Uncertainties on Dark Matter Direct Detection Results.}
\author[1]{Malcolm Fairbairn\thanks{E-mail address: malcolm.fairbairn@kcl.ac.uk}}
\author[1,2]{Tom Douce} 
\author[1]{Jace Swift}
\affil[1]{\em Physics, Kings College London, Strand, London WC2R 2LS, UK }
\affil[2]{\em Ecole Polytechnique, 91128 Palaiseau Cedex, France}

\begin{document}

\maketitle

\begin{abstract}
We attempt to estimate the uncertainty in the constraints on the spin independent dark matter-nucleon cross section due to our lack of knowledge of the dark matter phase space in the galaxy.  We fit the density of dark matter before investigating the possible solutions of the Jeans equation compatible with those fits in order to understand what velocity dispersions we might expect at the solar radius.  We take into account the possibility of non-Maxwellian velocity distributions and the possible presence of a dark disk.  Combining all these effects, we still find that the uncertainty in the interpretation of direct detection experiments for high ($>$100 GeV) mass dark matter candidates is less than an order of magnitude in cross section.
\end{abstract}

%\maketitle

\section{Introduction}
Many separate pieces of evidence exist pointing towards the existence of approximately 7 times as much dark matter as baryonic matter in the Universe, for example, galactic rotation curves \cite{rubin}, clusters of galaxies \cite{zwicky}, lensing studies \cite{lensing} and the growth of structure \cite{wmap}.

At the time of writing though, we still do not know what the precise particle nature of the dark matter is.  A class of leading candidates for dark matter are weakly interacting massive particles (WIMPs) which are charged under the standard model weak force.  These models are favoured because they lead to a thermal relic abundance roughly compatible with the observed density of dark matter required to explain astrophysical observations without the requirement of fine tuning of the dark matter mass.  

There are three ways of looking for WIMP dark matter particles, the first is to create them directly in colliders, and researchers at the LHC are working hard to do this \cite{lhcdm}.  The second way is to look for the standard model particles created when WIMP dark matter annihilates with itself in space, producing high energy gamma rays and cosmic rays, research into which is being spear-headed by the Fermi gamma ray telescope.   The third way is to look for the elastic collisions of dark matter particles with nuclei in purpose built underground detectors.  

In order for different dark matter direct detection experiments to compare constraints with each other they have to assume a particular halo model and the one they generally use is known as the Standard Halo Model (SHM).  This model is one of the simplest possible solutions of the Jeans equation but as we shall discuss in more detail, we do not expect the density profile or the velocity distribution of the dark matter in a real galaxy like our own to follow this and in fact there is an uncertainty in the local density and a large amount of uncertainty in the velocity distribution of dark matter.

There are many different analyses which have taken this into account - large combined analysis data packages contain various assumptions which take {\it some aspects} of this uncertainty into account.  Those uncertainties are usually only presented in the large combined analyses taking into account all the errors on other quantities including those from particle physics and constraining particular models of particle physics such as various versions of the constrained minimal supersymmetric standard model.  They don't necessarily present how the constraint on the cross section changes as a function of mass.  

We would like to estimate the error on the cross section in an independent way such that it can be understood and used by an independent researcher who comes up with a theory that makes a particular prediction for the dark matter-nucleon scattering cross section.

The dark matter direct detection experiment with the leading constraint on the dark matter nucleon cross section at the time of writing is the XENON-100 experiment, the constraint for dark matter particles of mass $M_{dm}\sim 200$ GeV being around $X\times 10^{-44}$cm$^2$ \cite{XENON100}.  This actually places constraints on a class of SUSY models known as the focus point.  Many collider phenomenology theorists are interested in a realistic estimate of what the errors are on this result, independent of a particle theoretical framework.  The experimental errors have already been quantified by the XENON team, so what remains is the uncertainty due to astrophysical unknowns.

First we will make the simplifying assumption that the bulk of the dark matter halo is spherical.  This neglects many aspects of the distribution of dark matter which we might expect to exist such as the oblate/prolate nature of the halo \cite{jesper} and axisymmetry \cite{GFBaxi}, although we don't expect these effects to change the conclusions a great deal since the constraints on the local dark matter density are mainly based on observations in the inner part of the galaxy where baryons will have increased the symmetry of the halo.  We will consider the possible presence and effect of a dark disk \cite{read}.

Unlike many previous analyses we also focus on the whole mass regime rather than concentrating on the low mass one.  There is a great deal of uncertainty in the event rate in detectors at very low mass ($M_{dm}<15 GeV$) because the uncertainties in the velocity distributions of dark matter become most pronounced in the high velocity tails, which are the only regions of phase space which allow a low mass dark matter particle to give rise a signal in direct detection experiments.  There are many papers attempting to explain the discrepancy between the DAMA annual modulation signal \cite{dama} and the XENON100 constraint, so we will not attempt to add to this debate on this occasion.  We are simply interested to understand what the latest constraints are really telling us in terms of numbers on the dark matter-nucleon cross section.

In section \ref{sectiondir} we will introduce the standard halo model and the equations for dark matter direct detection and reconstruct our version of the XENON100 constraint.  After that in section \ref{density} we will discuss the constraints and reconstruction of the density profile of the Milky Way Galaxy.  This section will be a poor mans reconstruction of the careful analysis of Catena and Ullio \cite{ulliocatena} with some small differences but we are not only interested in the density of dark matter in the solar neighborhood, but also the density profile as a function of radius as this affects the solution of the Jeans equation.  Section \ref{velocity} will address these solutions and explain why this adds another level of uncertainty to the problem as it changes the velocity dispersion in the solar system.  We will also discuss the effects of non-Gaussianity of the velocity distribution.

In section \ref{results} we will explain our procedure for combining the various astrophysical uncertainties and present results.  In section \ref{darkdisk} we investigate the effect a dark disk has on the uncertainty in the signal before finally in section \ref{conclusions} we will reflect upon our analysis and the results while suggesting possible future studies.

Such an estimate on the error associated with dark matter direct detection is almost subjective, since it involves estimating something that we cannot observe directly.  The particular uncertainties that any researcher will focus on might be slightly different (see e.g. \cite{mccabe}),  this is our best attempt to estimate these uncertainties.

\section{Direct Detection of Dark Matter\label{sectiondir}}

Since we expect dark matter to be moving with velocities of the order of $10^{-3}$c in the solar system, and for their masses to be around 10 GeV - 1 TeV the typical recoil velocity in a direct detection experiment will be in the keV energy range.

One can define a differential rate \cite{green,lewin}
\begin{equation}
\label{drde}
\frac{{\rm d} R}{{\rm d}E}(E) =
             \frac{\sigma 
             \rho}{2 \mu^2 m_{dm}}
             A^2 F^2(E)   \int_{|{\bf v}| \geq v_{{\rm min}}} 
            \frac{f({\bf v})}{v} {\rm d}^3 {\bf v}     \,, 
\end{equation}

where $E$ is the recoil energy, $\sigma$ is the dark matter nucleon cross section,  $m_{dm}$ is the dark matter mass, $\mu= (m_{\rm p} m_{dm})/(m_{{\rm p}}+ m_{{dm}})$,  $A$ is the atomic mass number of the target nuclei which enters due to the quantum mechanical coherence effect of all the nuclei lying within the Compton wavelength of the dark matter particle.  $F(E)$ is the nuclear form factor of the target nuclei and $\rho$ is the local density of dark matter.  Throughout this work we will use the Helm form factor, uncertainties in nuclear form factors can also change significantly the interpretation of dark matter direct detection experiments.  The  function $f({\bf v})$ is the (appropriately normalised) dark matter velocity distribution and $v_{min}$ is the minimum velocity which could create a recoil of energy $E$, which depends upon $M_{dm}$ and the mass of the nucleus rather than the proton.

The velocity distribution function $f({\bf v})$ is defined in the rest frame of the galaxy, relative to which we are moving at velocity $v_{circ}$ as we circle the centre of the galaxy at a couple of hundred kilometres per second (more discussion with regards to this precise value later).  $f({\bf v})$ is truncated at the escape velocity $v_{esc}$ corresponding to our position in the Milky Way.

The overall rate of events depends therefore not only on $\sigma$ and $m_{dm}$ as we would expect but also on $\rho$, $f({\bf v})$, $v_{circ}$ and $v_{esc}$.  We have limited information about these quantities based upon incomplete astrophysical data and incomplete theoretical knowledge in the form of both analytical and numerical solutions of the collisionless Boltzmann equation (i.e. idealised solutions of the Jean's equation and N-body simulations of galaxies).

In order to even begin to estimate the values of these parameters, we start by making some simplifying assumptions.  The first is that the phase space distribution of dark matter and the gravitational potential of the galaxy are both spherically symmetric in position space.  This is clearly far from a perfect assumption at small radii as we know that our galaxy is very much a disk, however we do not expect the dark matter to form a disk as readily as the baryonic component of the Galaxy since it cannot experience pressure or lose energy, so it is much less prone to forming viscous conglomerates which re-distribute angular momentum, like disks.  We will however later look at the possibility of a dark disk.

Dark matter direct detection experiments need to compare their results with each other and they therefore have an agreed industry standard known as the "Standard halo model''.  This model is an example of a spherically symmetric solution along with a number of further simplifying assumptions.  The first is that the halo has a density profile $\rho\propto r^{-2}$ which would lead to a constant rotational velocity if the gravitational potential of baryons were neglected.  The second is that the velocity distribution is precisely a Maxwell-Boltzmann distribution and is isotropic in all directions.  It can be verified that in order to solve equation (\ref{jeans}) in the light of these assumptions, the velocity dispersion $\sigma=v_{circ}/\sqrt{2}$ is a constant which does not change with radius and the standard value of $v_{circ}=220$kms$^{-1}$ is usually chosen.  This solution of the Jeans equation is known as the isothermal sphere for obvious reasons.  The only caprice the conservative dark matter experimentalists usually allow themselves is a variation of the escape velocity to see how sensitive their constraints are to this quantity which is not well defined for this solution. 

While we do not  know in detail the real distribution of dark matter in the Galaxy, we do not expect the isothermal sphere to be a good approximation.   N-body simulations of dark matter suggest that density profiles usually take the form of a profile which is indeed denser in the centre of the galaxy but the density rises more slowly at low radii and steeper at larger radii \cite{NFWein}.

We shall look at how we can model the density of dark matter in a more realistic way then we shall see what implications this has for the velocity dispersion of dark matter.  Having done that we will estimate how this would effect the event rate in direct detection experiments.

\section{Density Profile of Dark Matter\label{density}}

If the dark matter is distributed smoothly in the galaxy in a similar way to that suggested by N-body simulations, the density in the solar system is remarkably well constrained to be around $\rho\sim 0.2-0.4$ GeV cm$^{-3}$.  Recent studies agree with this \cite{deboer} and the most detailed study to date constrains the dark matter density very tightly \cite{ulliocatena}. 

\subsection{Substructure and Graininess}

In this work we will be assuming that the distribution of dark matter is smooth.  This is not obvious as it has been shown theoretically and in simulations that there are halos within halos down to objects which are roughly the size of the solar system \cite{stefan,moore}.  It has been estimated however that these substructures will be tidally disrupted as they pass through halos the size of the Milky Way such that one can view the density distribution as being essentially entirely smooth to one part in ten thousand \cite{schneider}.  This fits with beautiful results from N-body simulations where objects similar to caustics are clearly visible in phase space close to the virial radius of halos, but deep in the centre of dark matter halos (where we lie) so many streams are overlapping and so much tidal interaction has occurred that the density is essentially completely smooth \cite{vogelsberger}.

\subsection{The Einasto Profile}

In \cite{ulliocatena} the authors assume that the dark matter distribution is spherical, but take into account the non-sphericity of the baryonic disk.  They look at both Einasto, NFW and Burkert functions for the density profiles of dark matter.  In this work we will only use the Einasto profile since it is the one which seems to fit N-body simulations most closely and can give rise to relatively cuspy and cored profiles.  The Einasto profile takes the following form
\begin{equation}
\rho(r)=\rho_{-2}\exp\left[-\frac{2}{\alpha}\left(\left\{\frac{r}{r_{-2}}\right\}^\alpha-1\right)\right]
\end{equation}
and there are therefore three free parameters, the radius $r_{-2}$ at which the logarithmic derivative of density $\gamma=d\ln\rho/d\ln r$ is equal to $-2$ and the density at that radius $\rho_{-2}$.  The third parameter is the power $\alpha$ which sets how quickly $\gamma$ varies as a function of $r$.  We will consider two situations, the first is to allow all three parameters to vary independently in order to fit the data.  The second situation is when we will constrain the relationship between the parameters using data from cosmological simulations.

\subsection{Concentration parameter inspired by cosmological simulations}

This latter situation takes into account the fact that one expects smaller halos which form earlier in the history of the Universe to retain some memory of the density of the Universe when they dropped away from the Hubble expansion and collapsed.  Smaller halos therefore are expected to possess denser, more compact cores and the concentration parameter, namely the ratio between the virial radius of each halo $r_{vir}$ and the scale radius ($r_{-2}$ in the case of the Einasto profile) is a function of mass \cite{bullockconcentration}.  In particular we will follow the work of Duffy et al \cite{duffy} where they parametrise the concentration parameter $c$ in the following way
\begin{equation}
\frac{r_{vir}}{r_{-2}}=c=A\left(M/M_{pivot}\right)^B\left(1+z\right)^C
\label{duffy}
\end{equation}
where $z$ is redshift which we will set to zero and the parameters $A,B,C$ and $M_{pivot}$ are taken from reference \cite{duffy}.  We will look at both situations where the parameters of the Einasto profile are set to fulfill equation (\ref{duffy}) and when they are not.  The relationship in equation (\ref{duffy}) is anyway only meant to indicate the central value of the concentration parameter as a function of mass for a large ensemble of halos.  The shape of individual halos will depend upon their history.

\subsection{Baryonic contraction}

The role of baryons in affecting the distribution of dark matter in halos is as controversial as ever.  For a long time it has been known that adiabatic contraction of the dark matter halo occurs when baryons lose energy and sink into the centre of the galaxy, pulling in dark matter after them \cite{blumenthalcontraction}.  The importance of considering the shape of dark matter orbits was first pointed out in \cite{gnedin} and tested to be a much better approximation than using circular orbits in work such as \cite{jesper}.  This effect enhances the density of dark matter in the central region of the galaxy and makes the logarithmic density gradient $\gamma$ steeper throughout the region of the galaxy where baryons dominate the gravitational potential.   

More recently it has been shown that the non-adiabatic feedback which follows such contraction can eject dark mater from this central region and actually decrease the central density \cite{governato}.  While we expect this to be very important for indirect detection for the annihilation of dark matter at the centre of the galaxy, it will have a lesser effect upon the dark matter at the solar radius where such effects should be less pronounced -  for Milky Way size halos the authors of \cite{governato} found that dark matter is ejected to form a core of size $\sim 2$kpc, much smaller than the 8.5 kpc radius of the sun.  The precise shape of the mass profile in the galaxy will however effect the solutions of the Jeans equations that will determine the phase space distribution and direct detection signal in the solar system.  It is rather difficult to include adiabatic contraction in one of the Monte Carlo Markov Chains since they take significantly more time than any of the other algorithms involved in each step of the chain.  For this reason we note that we are not critically sensitive to the central density of dark matter in the Milky Way since baryons dominate the potential at small radii and point out that the Einasto profile contains the freedom to model quite different shaped halos.

\subsection{Fitting the gravitational potential}
In order to fit the baryonic and dark matter contributions to the gravitational potential, we again follow quite closely the analysis of \cite{ulliocatena} with some simplifications.  We model the baryonic disk of the galaxy as an exponential disk such that the surface density $\Sigma_*(r)=\Sigma_{*0}\exp{-r/r_{d}}$ with a cut-off at 15kpc following observations while we treat the galactic bulge as a point mass at the origin.

We then fit the terminal velocities in the disk \cite{terminalvelocity} which involves taking into account the non-sphericity of the baryonic distribution.  We fit the Solar Radius $R_{\odot}$ and the value of $\Sigma_{*}(r_\odot)$ and the total mass within 1.1kpc $\Sigma_{tot}(r_\odot)$.  We then fit the total mass within 50kpc and the total mass within 100kpc \cite{masses} as well as the SDSS data of reference \cite{xue}, where like \cite{ulliocatena} we allow the velocity anisotropy parameter $\beta_*$ of the stellar population in that study to vary.  We also use the same data as Ullio and Catena to fit the local disk density in baryonic matter and in total \cite{diskbaryon,disktotal}.

\begin{figure}[htp]
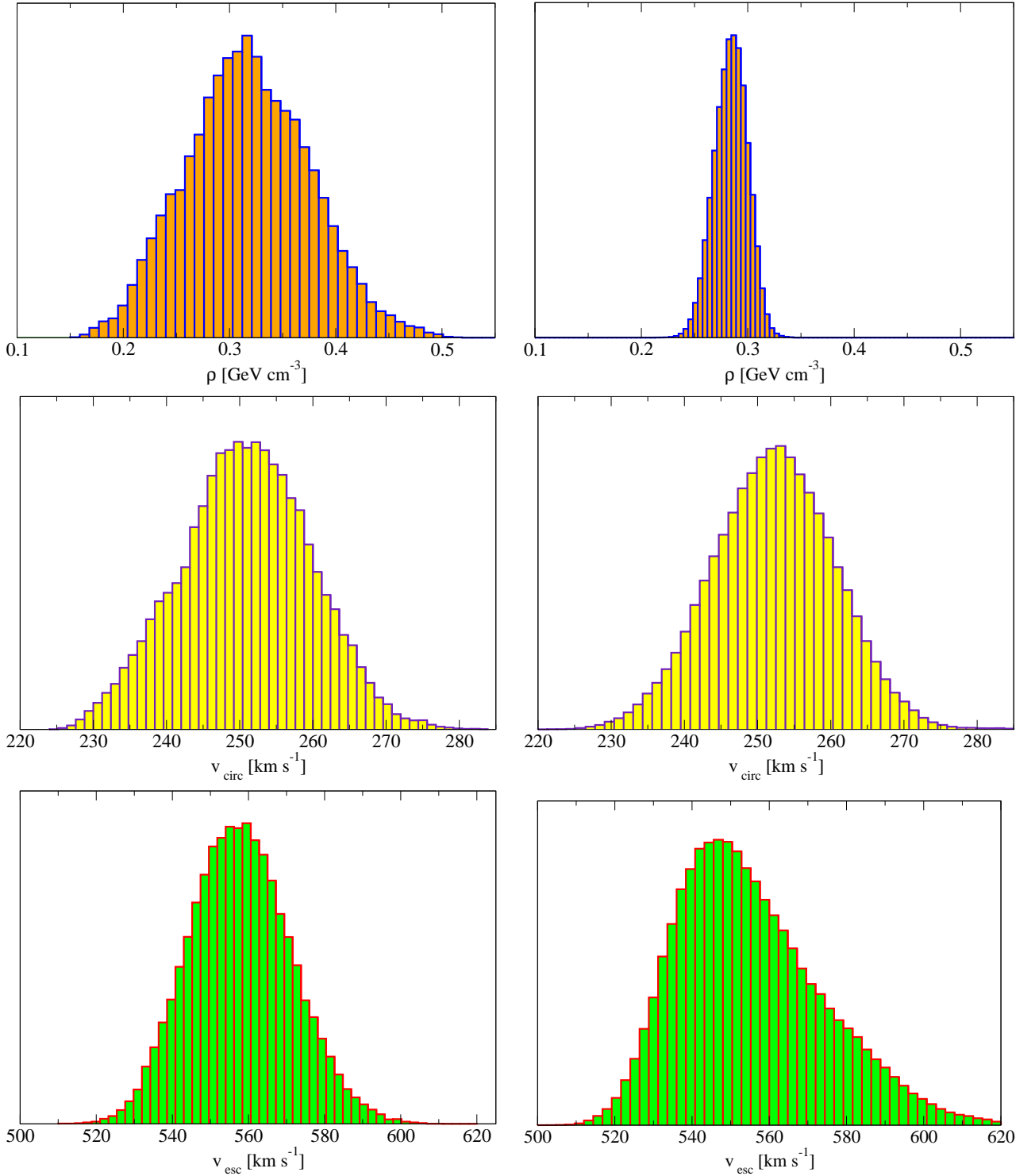

\begin{tabular}{cc}
\includegraphics[clip,width=0.5\columnwidth]{eindens.eps} & \includegraphics[clip,width=0.5\columnwidth]{wmapdens.eps}\\
\includegraphics[clip,width=0.5\columnwidth]{einvc.eps} & \includegraphics[clip,width=0.5\columnwidth]{wmapvc.eps}\\
\includegraphics[clip,width=0.5\columnwidth]{einvesc.eps} & \includegraphics[clip,width=0.5\columnwidth]{wmapvesc.eps}
\end{tabular}
\caption{\it Parameters for the Milky Way for the two situations mention in the text.  The top row is the local dark matter density in the solar system.  The middle row is the circular velocity of the Sun around the galaxy and the lower row is the escape velocity from the solar radius.  The plots on the left correspond to the situation where all three parameters of the Einasto profile are allowed to vary independently while the plots on the right correspond to those combinations of parameters which respect equation (\ref{duffy}) \label{compare1}}
\end{figure}

The results of the fit can be seen in figure \ref{compare1} and show reasonable agreement with other estimates in the literature (\cite{deboer, ulliocatena}).  Any differences might be explained by the fact that we don't use precisely the same data set as Ullio and Catena (we do not include the proper motions) and that our definition of the disk size is slightly different.  These results should not be taken too seriously on their own - the purpose of this paper is not to present a definitive estimate of the local dark matter density, but rather to estimate the magnitude of typical uncertainties in the direct detection signal.

\section{Velocity Distribution of Dark Matter\label{velocity}}

Once we have obtained a density of dark matter and baryons which fits the available data we need to look at what values of the velocity distribution are consistent with that density profile.  In order to do that we need to solve the collisionless Boltzmann equation, the moments of which are solutions to a differential equation known as the Jeans equation \cite{binneytremaine}

\subsection{Jeans Equation}
If we make the assumption of spherical symmetry for the distribution of mass, we can then write down the equation for the second moment of the collisionless Boltzmann equation for dark matter known as the Jeans equation
\begin{equation}
\frac{1}{\rho}\frac{\partial}{\partial r}\left(\rho\sigma_r^2\right)+\frac{2\beta\sigma_r^2}{r}=-\frac{GM(r)}{r^2}
\label{jeans}
\end{equation}
where $\rho$ is the density of dark matter as a function of radius, $\sigma_r$ is the dispersion of the component of the dark matter velocity parallel to the line between the point in question and the centre of the galaxy,  the velocity anisotropy parameter $\beta$ is
\begin{equation}
\beta=1-\frac{\sigma_t^2}{\sigma_{r}^2}
\end{equation}
where $\sigma_t$ is the velocity dispersion perpendicular to $\sigma_r$.  Of course this assumption of spherical symmetry does not take into account the baryonic disk which we would expect to change the velocity distribution of dark matter somewhat.  In particular one might expect that the disk would lead to an increase in the velocity dispersion in the $z$ direction perpendicular to the disk due to gravitational focusing (and a corresponding increase in density \cite{tremaine}).  However since in this work we are interested in the uncertainties in the expected signal rather than making a robust prediction of the signal itself we will continue by neglecting this asymmetry as solutions of the Jeans equation taking into account the shape of the disk potential would be significantly more complicated.

We cannot directly observe the velocity distribution of dark matter since we have not yet actually detected any dark matter.  At this stage therefore we move away from those quantities which can be fit by comparison with observations to those parameters which we cannot measure.  In particular the solution of the Jeans equation depends not only upon the gravitational mass as a function of radius which we can estimate and compare wit observations, but also upon the velocity anisotropy of dark matter which cannot be measured.  The only insight we have into the likely values of this parameter comes from theory, both analytical and numerical considerations.  A researcher who tries to look at the kind of velocity distributions chosen from random orbits which can fit a given density profile will quickly convince themselves that the degeneracies are too large to give any kind of indication of what the velocity profile should be as a function of radius.  we are therefore forced into the situation where the best we can do is choose some functional form for $\beta(r)$ which is not too restrictive while not having too many parameters.  We therefore assume the following ansatz for $\beta(r)$
\begin{equation}
\beta(r)=\frac{\beta_0+\beta_\infty(r/r_\beta)^\eta}{1+(r/r_b)^\eta}
\end{equation} 
where $\beta_0$ is the value of $\beta$ at the origin, $\beta_\infty$ is its value at large $r$, the parameter $r_b$ is the radius of switchover between these two regimes and the power $\eta$ determines the rate of switchover.  We cannot observe the velocity anisotropy parameter so we choose $\beta_0$ and $\beta_\infty$ randomly for each solution of the Jeans equation chosen using the weighting $\beta=1-x_1/x_2$ where $x_1$ and $x_2$ are both random numbers between one and zero.  We assume some kind of relationship between $r_b$ and the $r_{-2}$ parameter of the Einasto profile, but allow some give in this, in particular we say that
\begin{equation}
r_\beta=r_{-2}10^{1.5x-0.75}
\end{equation}
where again, $x$ is a random number between zero and one.  Finally we assume that $\eta$ is a random number between zero and four, with a flat distribution.

It would certainly be valid to argue that these choices are somewhat arbitrary, however they are chosen to try and mimic the output from N-body simulations to some degree, in particular the relationship between $r_\beta$ and $r_{-2}$.  In fact $\beta$ profiles as a function of radius can often be more complicated, however we hope that our choices will to some degree be a reasonable reflection of the uncertainties in the solution of the Jeans equation for a given mass profile.

\begin{figure}[htp]
\centering
\begin{tabular}{cc}
\includegraphics[clip,height=50mm]{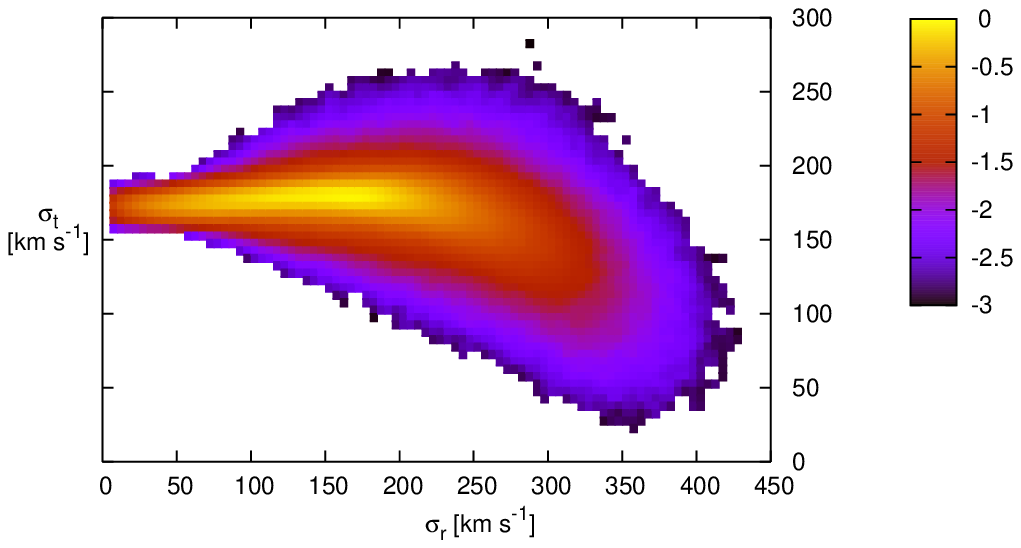} & \includegraphics[clip,height=50mm]{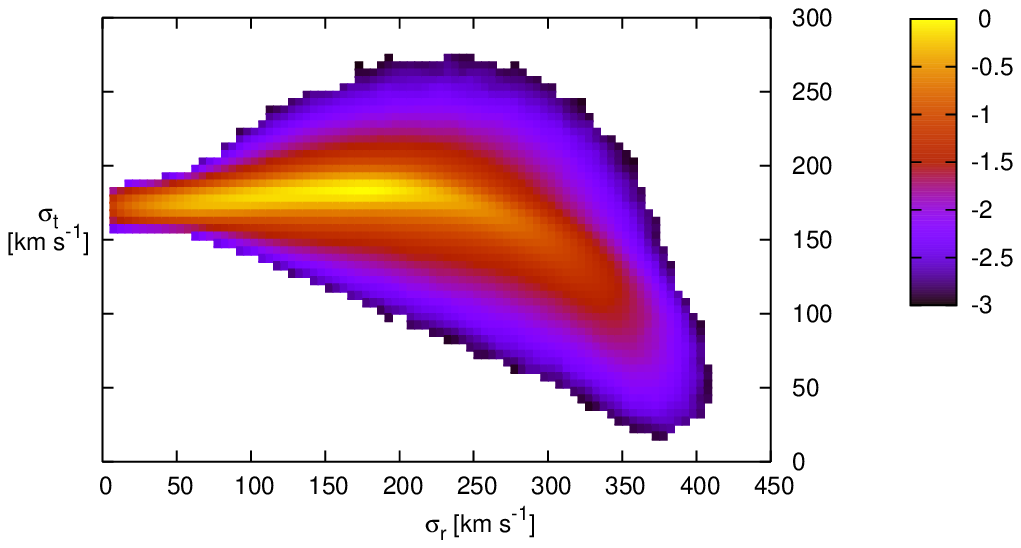}
\end{tabular}
\caption{\it The radial velocity dispersion $\sigma_r$ and the tangential velocity dispersion $\sigma_t$ at the solar radius as found by multiple solutions of the Jeans equation (\ref{jeans}).  Colours represent probability and the scale is logarithmic (i.e. the colour corresponding to -1 is 10\% of the maximum probability).}
\end{figure}

\subsection{Non Maxwellian Velocity Distributions.}

Once we have solved the Jeans equation, we have the second moment of the radial and tangential velocity distribution.  It is usually then assumed that this corresponds to the second moment of a Maxwellian (Gaussian) velocity distribution with a particular escape velocity.  However, this is not what is observed to happen in real simulations, where dark matter distributions are smooth, but are not generally speaking Gaussian in shape.

A moments thought shows that this is not a complete surprise.  The criteria for a Maxwell Boltzmann distribution is that the gas is in thermal equilibrium with itself and that it is an ideal gas (no finite size of particles or long range interactions).  While we expect the chemical potential and finite size effects to be irrelevant for dark matter in the kind of densities encountered in the solar system, there is clearly a long range potential involved, namely gravity.  Also the cross section for dark matter scattering off itself has been tightly constrained by observations such as the bullet cluster \cite{bulletselfinteraction} and will be irrelevant in thermalisation of dark matter.  The only thermalising force is therefore gravity, so we really shouldn't expect the dark matter to follow Gaussian distributions.

\begin{figure}[htp]
\begin{centering}
\includegraphics[clip,width=0.5\columnwidth]{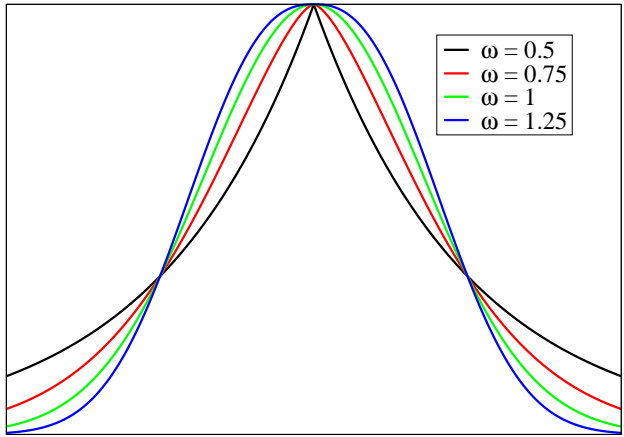} 
\caption{\it Schematic Figure Demonstrating the non-Maxwellian shape of different velocity profiles for different values of $\omega$ in equation (\ref{nong}).  Note $\omega=1$ is a Gaussian.\label{alphaschematicfig}}
\end{centering}
\end{figure}

Indeed, it has been known for a long time that dark matter simulations don't give rise to Maxwellian distributions \cite{nong}.  There are a number of ways one can model a non-Gaussian profile of velocities.  One well-motivated choice would be the Tsallis distribution \cite{hansenvergados, kuhlen} but in this work we consider instead the phenomenological distribution of reference \cite{fairbairnschwetz} where the velocity distributions follow the form
\begin{eqnarray}
\frac{1}{N_r}\exp\left[-\left(\frac{\tilde{v}_r^2}{f_r^2}\right)^{\omega_r}\right] \nonumber\\
\frac{2\pi v_t}{N_t}
\exp\left[{-\left(\frac{\tilde{v}_t^2}{f_t^2}\right)^{\omega_t}}\right]
\label{nong}
\end{eqnarray}
where $\omega_r$ and $\omega_t$ represent the non-Gaussianity in the radial and tangential direction respectively\footnote{we have changed the notation from reference \cite{fairbairnschwetz} from $\alpha$ to $\omega$ to avoid confusion with the $\alpha$ in the Einasto profile}.  These gives rise to profiles of non-Gaussian shape as can be seen in figure \ref{alphaschematicfig}.  
It was observed in reference \cite{fairbairnschwetz} that the parameter $\alpha$ is different in the $R$ and $T$ directions so following that work we assign random values to the two parameters in the following range: $0.5<\alpha_T<1$ and $0.75<\alpha_R<1.25$ which hopefully will parametrise our ignorance of the precise shape of the dark matter density profile appropriately.

\section{Calculating the Direct Detection Rate}
We have shown above the different factors affecting the density and velocity of dark matter in the solar system.  We will fit the density profile to the data using a Monte Carlo Markov chain and for each density profile added to the chain in this way we will then generate ten random velocity distributions. 

For each density and velocity distribution generated we therefore would like to calculate the corresponding rate in an experiment and for the purposes of this work we will look at the leading experiment at the time of writing, namely the XENON100 experiment \cite{XENON100}.  To obtain the constraints from equation (\ref{drde}) we first integrate it to obtain a cumulative rate as a function of energy and then we compare that with the cumulative rate in the XENON100 detector using the Kolmogorov-Smirnov test.  This is not quite what the XENON100 team did as they have their own Bayesian analysis chain, but for the purposes of estimating the errors due to uncertainties in astrophysical parameters it is sufficient.  Our simple analysis takes into account only the following:- the fiducial mass of the detector (48 kg) the exposure time (100.9 days) the energy range (8-44 keV) and the fact that the target is made of Xenon.  

We try to take into account the uncertainty in the relationship between observed and actual recoil energy due to our incomplete knowledge of the scintillation yield of Xenon, using the data presented in the XENON100 paper.  We do this in the following way - we transform the errors presented in figure 1 of reference \cite{XENON100} into errors on the energy for a given observed recoil, eight in total.  We then draw a random number from a Gaussian distribution and vary all the energies either up or down according to this energy varying distribution, which changes the $v_{min}$ in the calculation of $dR/dE$.  Again, this is designed to take into account the magnitude of this effect while we leave the rigorous statistics to the much more experienced XENON100 team.  Note however this approach assumes the errors are not strictly speaking independent and that the scintillation yield will be consistently higher or lower than the centre of the distribution, as opposed to oscillating or jumping around.  This seems like a reasonable assumption.

We also note that the XENON100 analysis does take into account one of the astrophysical uncertainties - namely the escape velocity and when we look at the uncertainty in the cross section for XENON100 we will take this into account, using the values they obtain from reference \cite{escape}, i.e. ($544^{+64}_{-46}$)km s$^{-1}$.

\begin{figure}[htp]
\centering
\begin{tabular}{cc}
\includegraphics[clip,width=0.5\columnwidth]{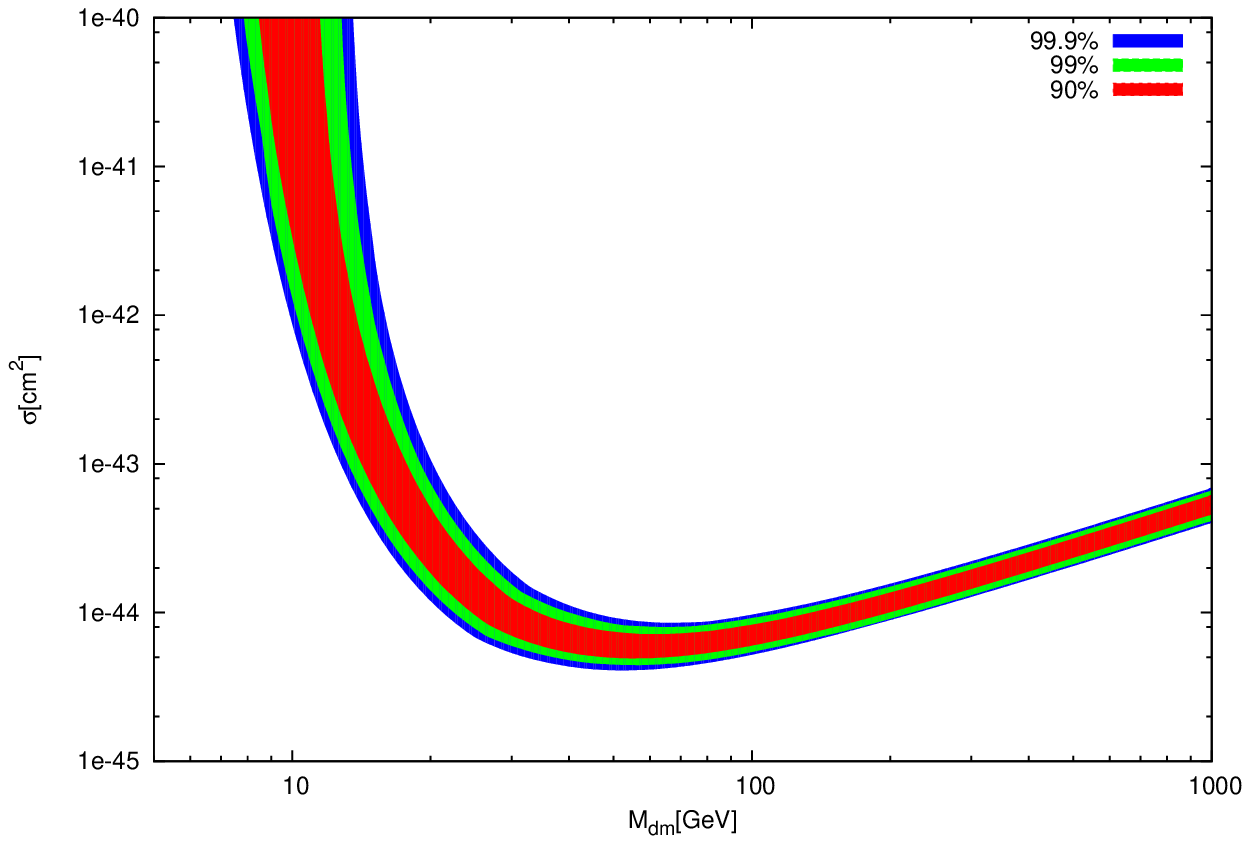} & \includegraphics[clip,width=0.5\columnwidth]{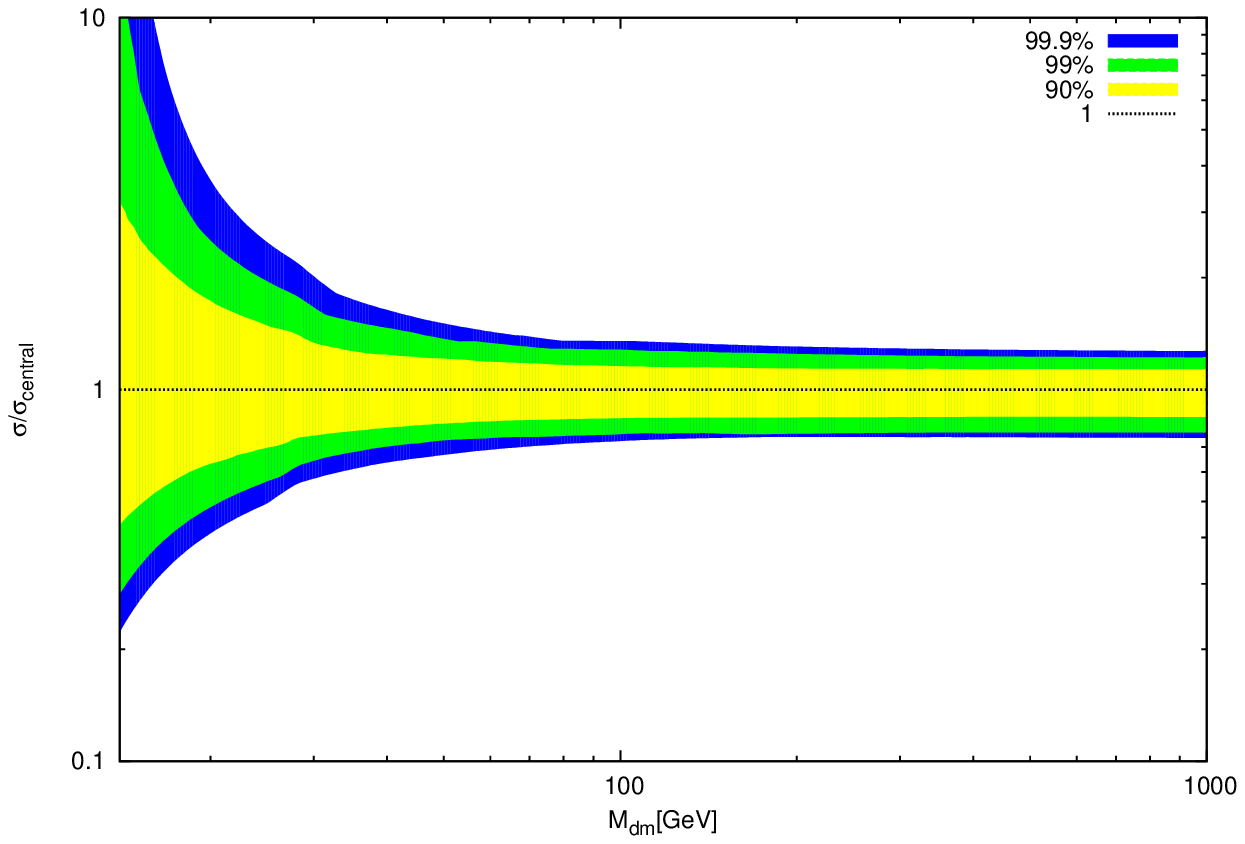}
\end{tabular}
\caption{\it The constraint we obtain in our simplified re-analysis of the results reported in the XENON100 paper \cite{XENON100}.  For each value of the escape velocity and for each assumption about the scintillation yield there is a different constraint on the cross section (90\% confidence level using Kolmogorow-Snirnov test).  Our constraints are in reasonable agreement with those of the XENON paper.  On the right we compare the variation in the constraint relative to the central value.  Note we only plot from about $M_{dm}=14$ GeV upwards as at lower dark matter masses, the spread dwarfs the uncertainties at higher masses and the constraints for low escape velocities disappear.\label{xenonplots}}
\end{figure}
\section{Results\label{results}}

In this section we will present our results.  First of all we will present our simple re-analysis of the XENON100 data and our estimate of the uncertainty in its constraint.

\subsection{XENON100 with the Standard Halo Model}

Figure \ref{xenonplots} shows the different constraints on the WIMP-nucleon cross section obtained in the standard halo model while allowing the escape velocity and the scintillation yield to vary as explained in the text.  The constraints are reasonably consistent with those presented in the XENON100 paper but are slightly different.  This is not surprising given the simplified nature of our analysis compared to that carried out by the experimental team.  Since this paper is not a re-analysis of the precise bound from XENON on the dark matter-nucleon cross section but rather the uncertainties introduced due to astrophysical uncertainties, this is acceptable.

Presumably the 90\% confidence level quoted in the XENON100 paper will correspond in spirit to the lower 90\% bound of the equivalent distribution found by their team. 

On the right hand side of figure \ref{xenonplots} we have plotted the relative spread in the XENON100 constraint on the cross section from uncertainties in the escape velocity and in the scintillation yield.  It is plain to see that at low energies, the uncertainty diverges.  We do not plot the region below about 12 GeV as for low escape velocities there is no signal at all capable of giving rise to a constraint from the XENON detector so the upper part of the distribution diverges to infinity.

\subsection{Our Spherical Halo Model}

\begin{figure}[htp]
\centering
\begin{tabular}{cc}
\includegraphics[clip,width=0.5\columnwidth]{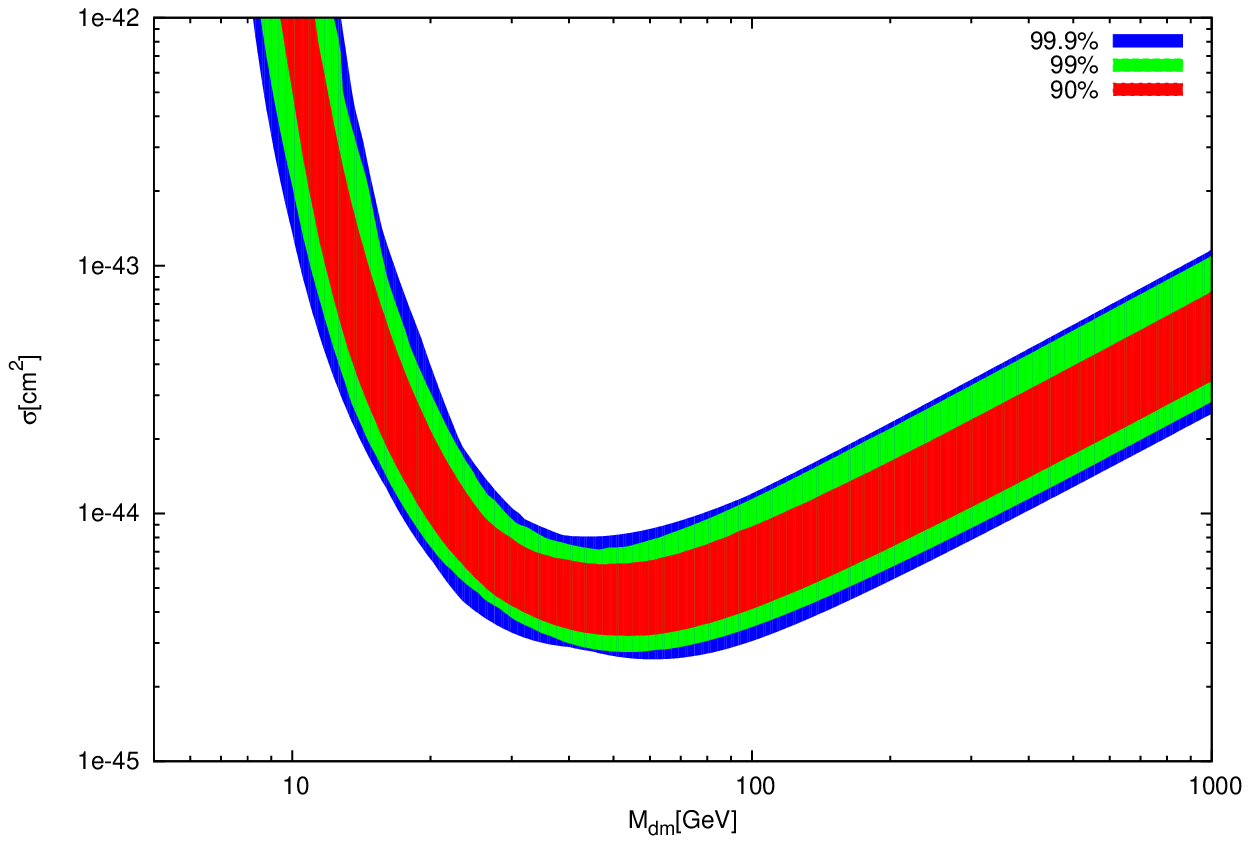} & \includegraphics[clip,width=0.5\columnwidth]{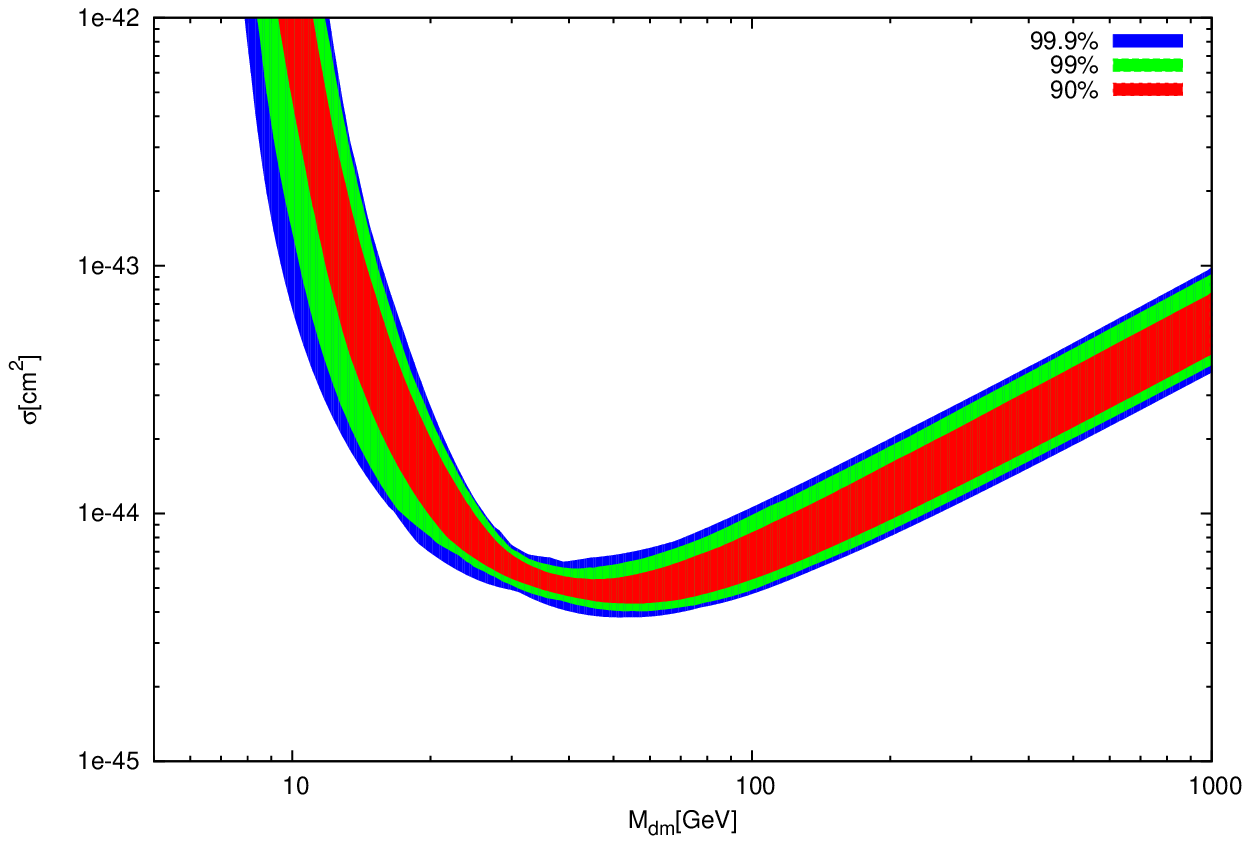}\\
\includegraphics[clip,width=0.5\columnwidth]{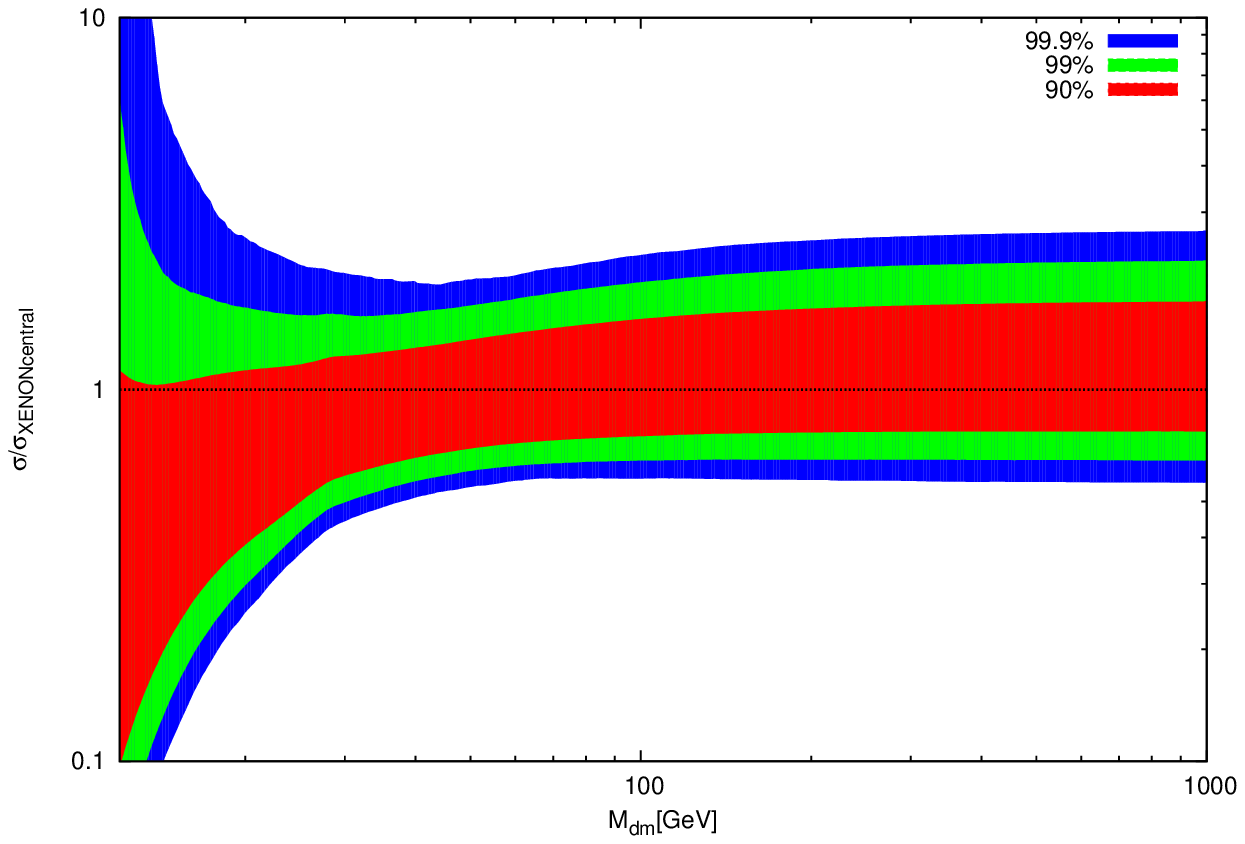} & \includegraphics[clip,width=0.5\columnwidth]{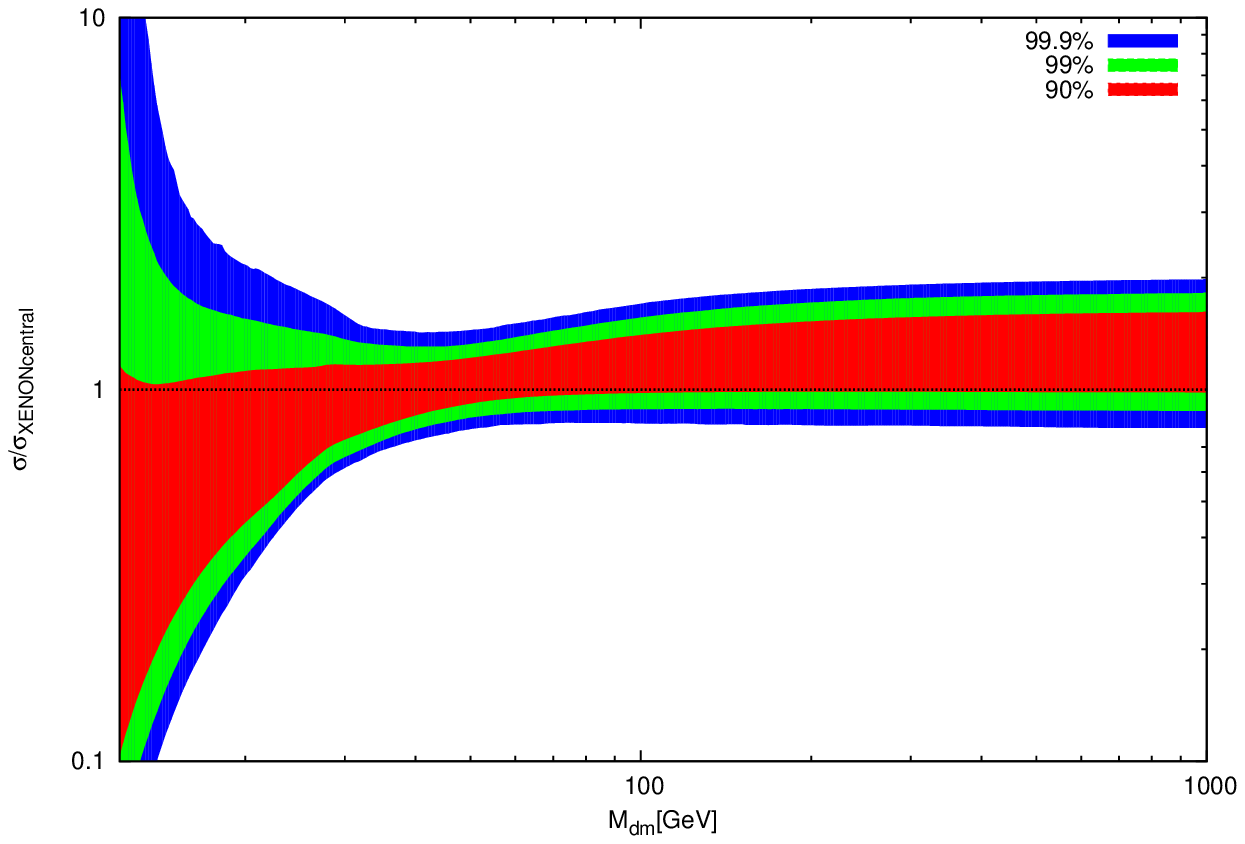}\\
\end{tabular}
\caption{\it Constraints on the dark matter nucleon cross-section for the more generalised spherical halo models considered in sections \ref{density} and \ref{velocity}.  On the left are the results when the three parameters in the Einasto Profile are allowed to vary while on the right is the case where the concentration parameter is fixed to respect the expectation from simulations.  Note that in this latter case, the density is very constrained, so the difference between the results on the right and the results in figure \ref{xenonplots} shows the magnitude of the variance in velocity dispersions upon solving the Jeans equation with a free anisotropy parameter.\label{spherical}}
\end{figure}

We calculated the exclusion limits for the different dark matter distributions obtained in section \ref{density} and \ref{velocity}. In figure \ref{spherical} we plot the constraints on the cross section that are obtained in this way.
We found that as expected the spread in the constraint on the cross section obtained for different dark matter masses varies quite a lot for different velocity and density distributions, but not a huge amount.  Although the change can be larger than 100\% the constraint on the cross section does not appear to vary more than a factor of a few at high masses.  This is quite surprising given the large amount that the velocity distribution changes, but for high mass wimps, the whole of the distribution is inside the region which can give rise to a signal, so the dependence on the velocity phase space is reduced.

The profiles where the density is constrained to respect the concentration parameter according to equation \cite{duffy} have less of a spread in density than the density profiles which are completely free to vary.  Because of this, the difference between the right hand side of the plots in figure \ref{spherical} and the standard halo model plots in figure \ref{xenonplots} are a good measure of the impact of allowing the velocity distributions freedom to vary upon the rate at direct detection experiments.

\section{The addition of a Dark Disk\label{darkdisk}}

\begin{figure}[htp]
\centering
\includegraphics[clip,width=80mm]{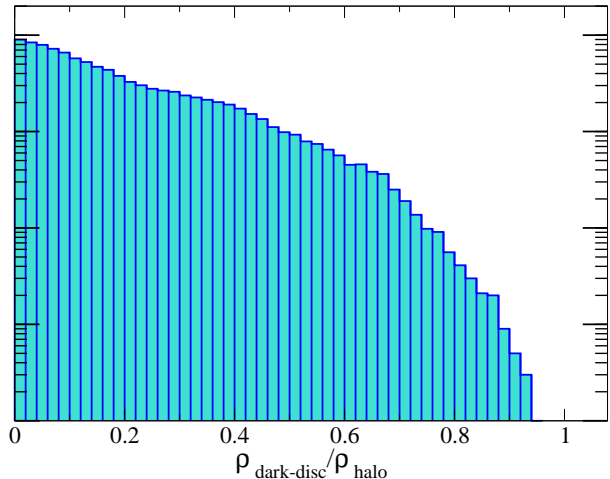}
\caption{\it Probability distribution of the radio of local dark matter which can exist in the form of a disk similar in radial profile to the baryonic disk with a total thickness of 2 kpc.\label{ddfrac}}
\end{figure}

A significant contribution to the direct detection signal could be due to the presence of a dark disk -  an enhancement in the density of dark matter which is aligned with the baryonic disk and either co-rotating or lagging behind \cite{read}.  We have tested this by doing the same fits to the density which we described in section \ref{density} but including an additional disk of dark matter with a thickness of 1.1kpc in order to estimate its effect upon the direct detection rate.  Clearly given the discrepancy between the overall surface density of the disk \cite{disktotal} and what can be seen in stars \cite{diskbaryon} there is some room for such a disk, although figure \ref{ddfrac} shows that it cannot be extremely significant relative to the local dark matter in the smooth spherical halo, we find it to be typically possible to find a disk some significant fraction of the local dark matter density, but always less than the local dark matter density.
The velocity distribution of this disk is also unknown, although it is clear that its velocity distribution will be less than the velocity distribution of the halo dark matter in order for it to remain bound to the disk \cite{infusin,fusin,fusinsims}.  We model the velocity distribution as being isotropic in the rest frame of the dark disk and of magnitude $50$km s$^{-1}<\sigma<120$km s$^{-1}$.  The lag behind the motion of the solar system we vary between zero and 150 km s$^{-1}$, both parameters chosen from a flat distribution.  We then calculate the event rate inside the detector for this case of a dark disk.
\begin{figure}[htp]
\centering
\includegraphics[clip,width=100mm]{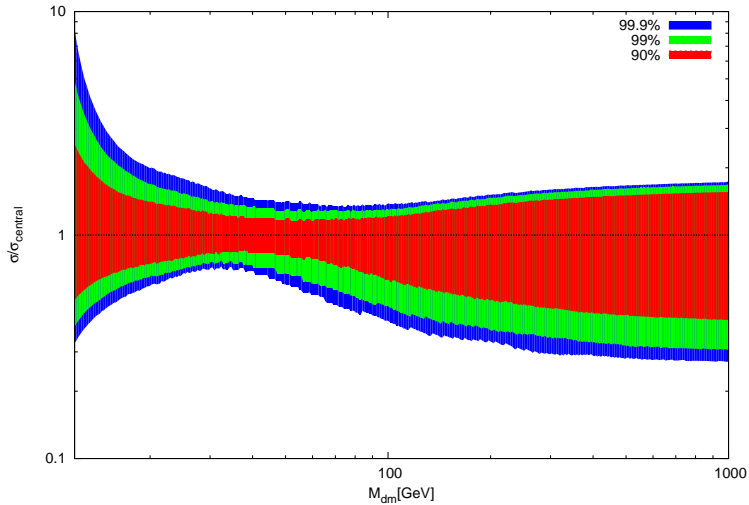}
\caption{\it The spread in the constraint on the cross section for halo models which contain a dark disk.  The density is found to vary relative to the halo density according to the distribution in figure \ref{ddfrac} while the velocities are varied as outlined in the text.  Even with a dark disk, the uncertainty in the constraint from direct detection experiments is still less than an order of magnitude for high mass dark matter candidates \label{ddspread}}
\end{figure}
The effect of this inclusion of a dark disk is shown in figure \ref{ddspread} which clearly shows that the effect of the dark disk is more pronounced at high masses, as shown in reference \cite{baudis}.  This is easy to understand as the velocity of the disk dark matter relative to the solar system will be much less than those velocities which can exist in the main body of the halo.

Still we do not find that the allowable density of dark disk is enough to radically change the results from dark matter direct detection experiments.  We therefore estimate the results presented by the XENON100 team are probably right to an accuracy within less than an order of magnitude.

\section{Conclusions\label{conclusions}}

There are many different collaborations currently designing, building and running dark matter direct detection experiments.  Unlike the energy and spatial distribution of, for example, neutrinos from the sun, dark matter is very far from being in thermal equilibrium, is non relativistic and owes its motion to the feeble gravitational force of all the matter in the local Universe as it makes its journey across the galaxy.  Following others, in this work we have tried to estimate the density of dark matter in the galaxy and then used that density to make estimates of the possible kinds of velocities dark matter could have in the solar system.

We have shown that the inclusion of uncertainties in density and velocity distributions both create an effect upon the rate of events in direct detection experiments and that these uncertainties are largest at lower velocities.  The purpose of this work is not to focus on the low mass regions but to investigate the errors due to astrophysical uncertainties upon searches for high mass dark matter particles.

We find that for a smooth spherical halo, taking into account all the possible uncertainties in the density distribution and the Jeans equation one can only increase the uncertainty in the bounds from direct detection experiments by a factor of a few.  In order to try and maximise the uncertainty, we have included the presence of a dark disk but have seen that given a very pronounced dark disk is incompatible with tracers of the local density of the Galaxy, this again can only increase the uncertainty by another factor of a few, leading to an overall uncertainty which is still less than an order of magnitude.

We believe that these estimates are pretty robust and that the only way they could be very wrong is if the current understanding of the smoothness of the local distribution of dark matter is wildly incorrect.  We haven't included uncertainties in nuclear form factors which would be an additional source of error but our conclusion is that the astrophysical uncertainty in the direct detection rate for high mass candidates is less than an order of magnitude.

\section*{Acknowledgments}
MF is grateful for funding provided by the UK Science and Technology Facilities Council.  This work was instigated by a question asked to MF by Graham Ross in December 2011 and is a complicated way of showing that the answer MF gave at the time was about right.

\end{document}